\documentstyle[12pt]{article}

\newcommand{\p}{\bot}

\newcommand{\dd}{\partial}

\newcommand{\de}{\delta}

\newcommand{\e}{\varepsilon}
\newcommand{\f}{\varphi}
\newcommand{\ls}{\left(}
\newcommand{\lks}{\left[}
\newcommand{\rs}{\right)}
\newcommand{\rks}{\right]}

\newcommand{\n}{\nu}
\newcommand{\m}{\mu}

\newcommand{\xx}{\vec x}

\newcommand{\str}[1]{\mathrel{\mathop{\longrightarrow}\limits_{#1}}}

\newcounter{form}
\newcommand{\dis}[1]{$$\displaylines{#1}$$}
\newcommand{\disn}[2]{$$\displaylines{\refstepcounter{form}
	    \label{#1} \hfill #2}$$}
\newcommand{\no}{\hfill \phantom{(\theform)}\cr \hfill}
\newcommand{\nom}{\hfill (\theform) \cr}

\newcommand{\reff}[1]{\ref{#1}}

\begin{document}

\title{Lattice Gauge Theories with polynomial Action \\
	and their canonical Formulation \\ on the Light Front}

\author{V. A. Franke\thanks{E-mail: franke@snoopy.phys.spbu.ru},
S. A. Paston\thanks{E-mail: Sergey.Paston@pobox.spbu.ru},
E. V. Prokhvatilov\thanks{E-mail: Evgeni.Prokhvat@pobox.spbu.ru}\\
St.-Peterspurg State University, Russia}

\date{26 January 1998}

\maketitle
\begin{abstract}

   A  lattice  gauge  theory  with  an  action   polynomial    in
independent field variables is considered. The link variables are
described by unconstrained complex matrices  instead  of  unitary
ones. A mechanism which permits to switch off in  the  continuous
limit the  arising  extra  fields  is  discussed.  The  Euclidean
version of the theory with an 4-dimensional lattice is described.
The canonical form of this theory  in  Lorentz  coordinates  with
continuous time  is  given.  The  canonical  formulation  in  the
light-front  coordinates  is  investigated  for  the  lattice  in
2-dimensional transverse space and for the 3-dimensional  lattice
including one of the light-like coordinates. The light-front zero
mode problem is considered in the  framework  of  this  canonical
formulation.

\end{abstract}

\newpage
\section{Introduction}

Ultroviolet regularization of nonabelian gauge theories via
introduction of space-time lattice is widely used
in nonperturbative considerations. It is usual to apply
Wilson-Polyakov lattice \cite{vilpol}, where gauge field is
described by unitary
matrices related to lattice links. This lattice is convinient
for numerical calculations owing to compactness of the  space of
parameters defining the unitary matrices. However the action of
the theory is nonpolynomial in this parameters which are independent
dynamical variables. This complicates the analitic investigation
of the theory. For example canonical formulation of lattice gauge
theory on the Light Front (LF) encounters difficulties
when Wilson-Polykov action is applied.
In this paper we consider the formulation
of gauge invariant lattice theory with an action polynomial in
independent variables. The idea is rather simple and originates
from Bardeen-Pearson paper \cite{bapir} devoted to gauge theory on the LF
with transverse space lattice. It can be explained as follows.

  Let us introduce a cubic lattice in space-time and relate
arbitrary complex $N\times N$ matrix $W$  to every link with positive
direction respectively to corresponding coordinate axis. For the same
link having opposite direction we use hermitian conjugated matrix
$W^+$. With any closed directed loop on the lattice one
relates
the trace of the product of such matrices taking into account the
direction of each link along the loop. Gauge transformations act
on this matrices like in Wilson-Polyakov lattice theory, i.e.
by $N\times N$ unitary matrices at the end points of corresponding
links. The above mentioned traces
belonging to closed loops remain invariant under this
transformations. One can use linear combinations of such invariants to
construct the action. In this theory the role of independent dynamical
variables play the matrix elements of $W$ and $W^+$,  and the
action is polynomial in this variables. After transition to continuous
space one gets a theory with usual gauge fields and with equal number
of additional (extra) fields. This extra fields can be made to acquire
infinite mass
in the limit of zero lattice spacing and to switch off from the theory.
Recently the analog of Bardeen-Pearson approach was applied to
$(2+1)$-dimensional gauge theory on the LF \cite{dal1,dal2}.
All extra fields was kept
in the theory in order to describ phenomenologically the effective interaction at
low energies \cite{dal2}. It was possible to reproduce the mass spectrum (known from
Wilson-Polyakov lattice calculations) by fitting free parameters
in effective interaction terms.

  The objective of our paper is to develope this approach in a
general form using appropriate choice of dynamical variables.
This chois allows to control easily the naive continuous limit
of the model and to illuminate the correspondence whith usual
continuous theory.
We start
with $U(N)$ gauge theory on 4-dimensional Euclidean lattice
and introduce the
gauge invariant "mass" term
for extra fields.
These fields decouple when the corresponding mass gos to
infinity in continuous limit.
To get $SU(N)$ gauge theory we add to the
action another "mass" term which gives "infinite mass" to abelian
part of the $U(N)$ field in the limit of zero lattice spacing.

Further we develope canonical formalism
in Lorenz coordinates using
3-dimensional space lattice and
in the LF coordinates using transverse space lattice
(here we apply periodic boundary
conditions for gauge fields defined on finite interval of light-like
coordinate
$x^-={1\over\sqrt{2}}(x^0-x^3)$ with the $x^+={1\over\sqrt{2}}(x^0+x^3)$ being the
"time"). We show the advantage of lattice LF formulation in solving
canonical zero mode problem \cite{fra,kal}.

The layout of the rest of the paper is as follows. In Sec.~2 it is
considered the lattice model in 4-dimensional Euclidean space with the action
polynomial in independent field variables. Here the cubic lattice is introduced
for all four dimesions. In Sec.~3 it is considered the gauge theory in
canonical form in Lorentz coordinates with continuous time and with
3-dimensional cubic lattice in the space coordinates. In Sec.~4 it is
considered the gauge theory in the LF coordinates. The Sec.~4.1 is devoted to
canonical formulation on the LF with continuous light-like coordinates and
with a lattice in two transverse coordinates. In Sec.~4.2 it is considered
briefly the possibility to introduce  a lattice simultanously
in transverse and in
light-like $x^-$-coordinates.

\section{Gauge field theory on the lattice
in 4-dimensional Euclidean space}

   We  consider  $U(N)$ gauge  theory  without  scalar  and  spinor
fields. We introduce 4-dimensional cubic lattice  and  denote  by
$e_\mu$ the vector  connecting  two  neighbouring  sites  on  the
lattice and directed along positive $\mu$- coordinate axis. The $x^0,
x^1, x^2, x^3$ denote the coordinates of the
sites, and the $a$ is  lattice  spacing.  We  relate
complex $N\times N$ matrix $W$ to a link directed from the site $x-e_\mu$
to the site $x$
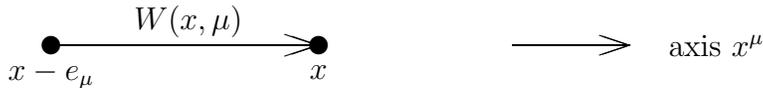
 \begin{figure}[h]
\vskip 8mm
\unitlength 1.00mm
\linethickness{0.4pt}
\begin{picture}(105.48,12.16)
\put(23.41,10.03){\line(1,0){35.56}}
\put(84.81,10.03){\line(1,0){15.50}}
\put(23.41,6.08){\makebox(0,0)[cc]{$x-e_\m$}}
\put(58.97,6.38){\makebox(0,0)[cc]{$x$}}
\put(41.65,12.16){\makebox(0,0)[cb]{$W(x,\m)$}}
\put(105.48,10.03){\makebox(0,0)[lc]{axis $x^\m$}}
\put(23.41,10.03){\circle*{2.50}}
\put(58.97,10.03){\circle*{2.50}}
\put(58.97,10.03){\line(-3,1){4.5}}
\put(58.97,10.03){\line(-3,-1){4.5}}
\put(100.31,10.03){\line(-3,1){4.0}}
\put(100.31,10.03){\line(-3,-1){4.0}}
\end{picture}
\end{figure}

\noindent
an the $W^+$ to the same link in opposite direction
 
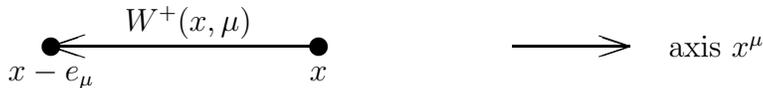
\begin{figure}[h]
\vskip 8mm
\unitlength 1.00mm
\linethickness{0.4pt}
\begin{picture}(105.48,12.16)
\put(23.41,10.03){\line(1,0){35.56}}
\put(84.81,10.03){\line(1,0){15.50}}
\put(23.41,6.08){\makebox(0,0)[cc]{$x-e_\m$}}
\put(58.97,6.38){\makebox(0,0)[cc]{$x$}}
\put(41.65,12.16){\makebox(0,0)[cb]{$W^+(x,\m)$}}
\put(105.48,10.03){\makebox(0,0)[lc]{axis $x^\m$}}
\put(23.41,10.03){\circle*{2.50}}
\put(58.97,10.03){\circle*{2.50}}
\put(23.41,10.03){\line(3,1){4.5}}
\put(23.41,10.03){\line(3,-1){4.5}}
\put(100.31,10.03){\line(-3,1){4.0}}
\put(100.31,10.03){\line(-3,-1){4.0}}
\end{picture}
\end{figure}

The elements of these matrices are considered as dynamical variables. We relate to any
closed  directed loop on the lattice the trace of the product of such
matrices seating on the links directed along this loop. For example,
the expression
 \dis{
{\rm Tr\;}\left\{W(x,\n)W(x-e_{\n},\m)W^+(x-e_{\m},\n)W^+(x,\m)\right\}.
}
is related to the loop shown in fig.~1.
 \begin{figure}
\vskip 8mm
\unitlength 1.00mm
\linethickness{0.4pt}
\begin{picture}(133.97,53.03)
\put(40.99,10.03){\line(1,0){35.56}}
\put(113.30,17.61){\line(1,0){15.50}}
\put(133.97,17.61){\makebox(0,0)[lc]{axis $x^\m$}}
\put(40.99,10.03){\circle*{2.50}}
\put(76.55,10.03){\circle*{2.50}}
\put(76.55,10.03){\line(-3,1){4.50}}
\put(76.55,10.03){\line(-3,-1){4.50}}
\put(128.80,17.61){\line(-3,1){4.00}}
\put(128.80,17.61){\line(-3,-1){4.00}}
\put(76.64,10.08){\line(0,1){35.56}}
\put(76.64,45.64){\line(-1,-3){1.50}}
\put(76.64,45.64){\line(1,-3){1.50}}
\put(40.94,45.68){\line(0,-1){35.56}}
\put(40.94,10.12){\line(1,3){1.50}}
\put(40.94,10.12){\line(-1,3){1.50}}
\put(76.59,45.73){\line(-1,0){35.56}}
\put(76.59,45.73){\circle*{2.50}}
\put(41.03,45.73){\circle*{2.50}}
\put(41.03,45.73){\line(3,-1){4.50}}
\put(41.03,45.73){\line(3,1){4.50}}
\put(80.31,49.70){\makebox(0,0)[lb]{$x$}}
\put(36.37,50.30){\makebox(0,0)[rb]{$x-e_{\m}$}}
\put(36.06,5.76){\makebox(0,0)[rt]{$x-e_{\m}-e_{\n}$}}
\put(80.31,6.06){\makebox(0,0)[lt]{$x-e_{\n}$}}
\put(83.03,27.58){\makebox(0,0)[lc]{$W(x,\n)$}}
\put(33.64,27.88){\makebox(0,0)[rc]{$W^+(x-e_{\m},\n)$}}
\put(59.70,53.03){\makebox(0,0)[ct]{$W^+(x,\m)$}}
\put(59.40,3.03){\makebox(0,0)[cb]{$W(x-e_{\n},\m)$}}
\put(109.07,21.78){\line(0,1){15.50}}
\put(109.07,37.28){\line(-1,-3){1.33}}
\put(109.07,37.28){\line(1,-3){1.33}}
\put(111.82,39.39){\makebox(0,0)[lb]{axis $x^{\n}$}}
\end{picture}
\caption{}
\end{figure}
 \begin{figure}
\vskip 8mm
\unitlength 1.00mm
\linethickness{0.4pt}
\begin{picture}(100.18,26.06)
\put(53.72,15.47){\makebox(0,0)[rc]{$x-e_\m$}}
\put(100.18,15.47){\makebox(0,0)[lc]{$x$}}
\put(58.87,15.79){\circle*{2.50}}
\put(94.43,15.79){\circle*{2.50}}
\put(79.27,20.98){\line(-3,1){4.50}}
\put(79.27,20.98){\line(-3,-1){4.50}}
\multiput(58.79,15.76)(0.20,0.12){10}{\line(1,0){0.20}}
\multiput(60.83,16.94)(0.23,0.11){9}{\line(1,0){0.23}}
\multiput(62.89,17.97)(0.26,0.11){8}{\line(1,0){0.26}}
\multiput(64.97,18.84)(0.35,0.12){6}{\line(1,0){0.35}}
\multiput(67.07,19.56)(0.42,0.11){5}{\line(1,0){0.42}}
\multiput(69.18,20.13)(0.53,0.10){4}{\line(1,0){0.53}}
\multiput(71.31,20.54)(0.72,0.09){3}{\line(1,0){0.72}}
\put(73.46,20.80){\line(1,0){2.17}}
\put(75.63,20.91){\line(1,0){2.19}}
\multiput(77.82,20.86)(1.10,-0.10){2}{\line(1,0){1.10}}
\multiput(80.02,20.66)(0.74,-0.12){3}{\line(1,0){0.74}}
\multiput(82.24,20.31)(0.45,-0.10){5}{\line(1,0){0.45}}
\multiput(84.48,19.80)(0.38,-0.11){6}{\line(1,0){0.38}}
\multiput(86.74,19.15)(0.33,-0.12){7}{\line(1,0){0.33}}
\multiput(89.01,18.33)(0.25,-0.11){9}{\line(1,0){0.25}}
\multiput(91.31,17.37)(0.23,-0.11){14}{\line(1,0){0.23}}
\multiput(58.79,15.76)(0.20,-0.12){10}{\line(1,0){0.20}}
\multiput(60.83,14.58)(0.23,-0.11){9}{\line(1,0){0.23}}
\multiput(62.89,13.55)(0.26,-0.11){8}{\line(1,0){0.26}}
\multiput(64.97,12.68)(0.30,-0.10){7}{\line(1,0){0.30}}
\multiput(67.07,11.96)(0.42,-0.11){5}{\line(1,0){0.42}}
\multiput(69.18,11.39)(0.53,-0.10){4}{\line(1,0){0.53}}
\multiput(71.31,10.98)(0.72,-0.09){3}{\line(1,0){0.72}}
\put(73.46,10.72){\line(1,0){2.17}}
\put(75.63,10.61){\line(1,0){2.19}}
\multiput(77.82,10.65)(1.10,0.10){2}{\line(1,0){1.10}}
\multiput(80.02,10.85)(0.74,0.12){3}{\line(1,0){0.74}}
\multiput(82.24,11.21)(0.45,0.10){5}{\line(1,0){0.45}}
\multiput(84.48,11.71)(0.38,0.11){6}{\line(1,0){0.38}}
\multiput(86.74,12.37)(0.33,0.12){7}{\line(1,0){0.33}}
\multiput(89.01,13.18)(0.25,0.11){9}{\line(1,0){0.25}}
\multiput(91.31,14.15)(0.23,0.11){14}{\line(1,0){0.23}}
\put(73.95,10.66){\line(3,-1){4.50}}
\put(73.95,10.66){\line(3,1){4.50}}
\end{picture}
\caption{}
\end{figure}

It should be noticed that the trace related to closed loop consisting
of the same links passed in both directions is not  identically
unity because the matrices $W$
are not unitary. (See, for example, fig.~2.)
The unitary matrices $U(x)$ of gauge transformations act on the $W$
and $W^+$ in the following way:
 \disn{1a}{
W(x,\m)\to W'(x,\m)=U(x)W(x,\m)U^+(x-e_{\m}),
\nom}
 \disn{1b}{
W^+(x,\m)\to W'^+(x,\m)=U(x-e_{\m})W(x,\m)U^+(x).
\nom}

Quntities related to closed loops on the lattice
as described are invariant under
this transformations.
In order to construct gauge invariant action having correct naive limit at
$a \to 0$ we use the analogy with formulation in continuous
space. We write
 \disn{2a}{
W(x,\m)=I-gaV_{\m}(x),
\nom}
 \disn{2b}{
W^+(x,\m)=I-gaV^+_{\m}(x),
\nom}
where the $g$ is the coupling constant.  Further, we define
in fundamental representation the generators
$T^a$ of $SU(N)$-groop and the generator $T^0$ of $U(1)$ group.
The indeces like $a, b, \dots$ are related only to $SU(N)$ generators and the
indeces like $\tilde a, \tilde b, \dots$ run over all set of $U(N)$-generators.
We define them as follows:
 \disn{4a}{
T^{\tilde a}=T^{\tilde a +},
\nom}
 \disn{4b}{
{\rm Tr\;}(T^{\tilde a}T^{\tilde b})=\de^{\tilde a\tilde b},
\nom}
 \disn{4c}{
[T^a,T^b]=it^{abc}T^c,
\nom}
 \disn{4d}{
{\rm Tr\;} T^a=0,\qquad {\rm Tr\;} T^0=\sqrt{N}.
\nom}
We decompose the matrices $V_{\m}(x)$ and $V^+_{\m}(x)$ putting
 \disn{5a}{
V_{\m}(x)=B_{\m}(x)+iA_{\m}(x),
\nom}
 \disn{5b}{
V^+_{\m}(x)=B_{\m}(x)-iA_{\m}(x),
\nom}
where $B^+_{\m}=B_{\m}$, $A_{\m}^+=A_{\m}$,
and separate the abelian parts:
 \disn{6a}{
B_{\m}(x)=T^0b_{\m}(x)+\tilde B_{\m}(x),
\nom}
 \disn{6b}{
A_{\m}(x)=T^0a_{\m}(x)+\tilde A_{\m}(x),
\nom}
 \dis{
\tilde B_{\m}(x)=T^a B^a_{\m}(x),\quad \tilde A_{\m}(x)=T^a A^a_{\m}(x),\quad
{\rm Tr\;}\tilde B_{\m}=0,\quad {\rm Tr\;}\tilde A_{\m}=0.
}
In naive  limit  $a\to  0$  the  $\tilde A_{\m}$  coincides  with  usual
nonabelian $SU(N)$ gauge  field,  the  $a_{\m}$  becomes  abelian
gauge field, and $b_{\m}$,  $\tilde B_{\m}$  become  extra  fields  which
should be switched off.

  We can define an analog of covariant derivative.  For any field
$\f$ localized at the sites of the lattice it is
 \disn{10}{
\tilde{D}\f(x)={1\over a}\ls\f(x)-W(x,\m)\f(x-e_{\m})\rs=
\tilde{\dd}\f(x)+gV_{\m}(x)\f(x-e_{\m}),
\nom}
where
 \disn{8}{
\tilde{\dd}\f(x)={1\over a}\ls\f(x)-\f(x-e_{\m})\rs
\nom}
is the analog of usual derivative. It  follows  that  under  gauge
transformation
 \disn{11}{
\tilde{D}\f(x)\to \tilde{D'}\f'(x)=U(x)\tilde{D}\f(x),
\nom}
where $\f(x)\to \f'(x)=u(x)\f(x)$.

Hence,
 \disn{12}{
\ls\tilde{D}'_{\m}\tilde{D}'_{\n}-
\tilde{D}'_{\n}\tilde{D}'_{\m}\rs \f'(x)=
U(x)\ls\tilde{D}_{\m}\tilde{D}_{\n}-
\tilde{D}_{\n}\tilde{D}_{\m}\rs\f(x-e_{\m}-e_{\n}),
\nom}
and one can define the analog $G_{\m\n}(x)$ of the usual tensor field
wia the relation
 \disn{13}{
\ls\tilde{D}'_{\m}\tilde{D}'_{\n}-
\tilde{D}'_{\n}\tilde{D}'_{\m}\rs \f'(x)=
gG_{\m\n}(x)\f(x-e_{\m}-e_{\n}),
\nom}
where
 \disn{14}{
G_{\m\n}(x)=\tilde{\dd}_{\m}V_{\n}(x)-\tilde{\dd}_{\n}V_{\m}(x)+
g\ls V_{\m}(x) V_{\n}(x-e_{\m})-V_{\n}(x) V_{\m}(x-e_{\n})\rs=\no
={1\over a^2g}\ls W(x,\m)W(x-e_{\m},\n)-W(x,\n)W(x-e_{\n},\m)\rs.
\nom}
It can be represented in the form
\vskip -8mm
 \disn{r5}{
G_{\m\n}(x)={1\over a^2g}\Biggl(
\unitlength 1.00mm
\linethickness{0.4pt}
\begin{picture}(21.50,14.00)(2.00,7.20)
\put(10.00,4.00){\line(0,1){9.00}}
\multiput(10.00,13.00)(0.12,-0.35){6}{\line(0,-1){0.35}}
\multiput(10.00,13.00)(-0.12,-0.35){6}{\line(0,-1){0.35}}
\put(10.00,4.00){\circle*{1.00}}
\put(10.00,13.00){\circle*{1.00}}
\put(10.00,13.00){\line(1,0){9.00}}
\multiput(19.00,13.00)(-0.35,0.12){6}{\line(-1,0){0.35}}
\multiput(19.00,13.00)(-0.35,-0.12){6}{\line(-1,0){0.35}}
\put(19.00,13.00){\circle*{1.00}}
\put(20.00,14.00){\makebox(0,0)[lb]{\scriptsize $x$}}
\put(10.00,3.00){\makebox(0,0)[ct]{\scriptsize $x-e_{\m}-e_{\n}$}}
\put(10.00,14.00){\makebox(0,0)[cb]{\scriptsize $x-e_{\m}$}}
\end{picture}
 -
\unitlength 1.00mm
\linethickness{0.4pt}
\begin{picture}(28.00,13.00)(1.00,7.20)
\put(10.00,4.00){\circle*{1.00}}
\put(19.00,4.00){\circle*{1.00}}
\put(19.00,13.00){\circle*{1.00}}
\put(10.00,4.00){\line(1,0){9.00}}
\multiput(19.00,13.00)(0.12,-0.35){6}{\line(0,-1){0.35}}
\multiput(19.00,13.00)(-0.12,-0.35){6}{\line(0,-1){0.35}}
\put(19.00,4.00){\line(0,1){9.00}}
\multiput(19.00,4.00)(-0.35,0.12){6}{\line(-1,0){0.35}}
\multiput(19.00,4.00)(-0.35,-0.12){6}{\line(-1,0){0.35}}
\put(20.00,14.00){\makebox(0,0)[lb]{\scriptsize $x$}}
\put(10.00,3.00){\makebox(0,0)[ct]{\scriptsize $x-e_{\m}-e_{\n}$}}
\put(20.00,4.00){\makebox(0,0)[lt]{\scriptsize $x-e_{\n}$}}
\end{picture}
 \Biggr).
\nom}
\vskip 3mm
This field transforms under gauge transformations as follows:
 \disn{15}{
G_{\m\n}(x)\to G'_{\m\n}(x)=U^+(x)G_{\m\n}(x)U(x-e_{\m}-e_{\n}).
\nom}

Therefore the action
 \disn{r6}{
S_1={a^4 \over 4}\sum\limits_{x,\m,\n} {\rm Tr}\ls G^+_{\m\n}(x)G_{\m\n}(x)\rs
\nom}
is $U(N)$ gauge ivariant.  It can be represented as follows:
 \dis{
S_1={1\over 4g^2}\sum\limits_{x,\m,\n}{\rm Tr\,}\Biggl\{\Biggl(
\unitlength 1.00mm
\linethickness{0.4pt}
\begin{picture}(21.50,14.00)(2.00,7.20)
\put(10.00,4.00){\line(0,1){9.00}}
\multiput(10.00,4.00)(0.12,0.35){6}{\line(0,1){0.35}}
\multiput(10.00,4.00)(-0.12,0.35){6}{\line(0,1){0.35}}
\put(10.00,4.00){\circle*{1.00}}
\put(10.00,13.00){\circle*{1.00}}
\put(10.00,13.00){\line(1,0){9.00}}
\multiput(10.00,13.00)(0.35,0.12){6}{\line(1,0){0.35}}
\multiput(10.00,13.00)(0.35,-0.12){6}{\line(1,0){0.35}}
\put(19.00,13.00){\circle*{1.00}}
\put(20.00,14.00){\makebox(0,0)[lb]{\scriptsize $x$}}
\put(10.00,3.00){\makebox(0,0)[ct]{\scriptsize $x-e_{\m}-e_{\n}$}}
\put(10.00,14.00){\makebox(0,0)[cb]{\scriptsize $x-e_{\m}$}}
\end{picture}
 \hskip -1.5mm - \hskip -1.5mm
\unitlength 1.00mm
\linethickness{0.4pt}
\begin{picture}(28.00,13.00)(1.00,7.20)
\put(10.00,4.00){\circle*{1.00}}
\put(19.00,4.00){\circle*{1.00}}
\put(19.00,13.00){\circle*{1.00}}
\put(10.00,4.00){\line(1,0){9.00}}
\multiput(19.00,4.00)(0.12,0.35){6}{\line(0,1){0.35}}
\multiput(19.00,4.00)(-0.12,0.35){6}{\line(0,1){0.35}}
\put(19.00,4.00){\line(0,1){9.00}}
\multiput(10.00,4.00)(0.35,0.12){6}{\line(1,0){0.35}}
\multiput(10.00,4.00)(0.35,-0.12){6}{\line(1,0){0.35}}
\put(20.00,14.00){\makebox(0,0)[lb]{\scriptsize $x$}}
\put(10.00,3.00){\makebox(0,0)[ct]{\scriptsize $x-e_{\m}-e_{\n}$}}
\put(20.00,4.00){\makebox(0,0)[lt]{\scriptsize $x-e_{\n}$}}
\end{picture}
 \Biggr)\Biggl(
\unitlength 1.00mm
\linethickness{0.4pt}
\begin{picture}(21.50,14.00)(2.00,7.20)
\put(10.00,4.00){\line(0,1){9.00}}
\multiput(10.00,13.00)(0.12,-0.35){6}{\line(0,-1){0.35}}
\multiput(10.00,13.00)(-0.12,-0.35){6}{\line(0,-1){0.35}}
\put(10.00,4.00){\circle*{1.00}}
\put(10.00,13.00){\circle*{1.00}}
\put(10.00,13.00){\line(1,0){9.00}}
\multiput(19.00,13.00)(-0.35,0.12){6}{\line(-1,0){0.35}}
\multiput(19.00,13.00)(-0.35,-0.12){6}{\line(-1,0){0.35}}
\put(19.00,13.00){\circle*{1.00}}
\put(20.00,14.00){\makebox(0,0)[lb]{\scriptsize $x$}}
\put(10.00,3.00){\makebox(0,0)[ct]{\scriptsize $x-e_{\m}-e_{\n}$}}
\put(10.00,14.00){\makebox(0,0)[cb]{\scriptsize $x-e_{\m}$}}
\end{picture}
 \hskip -1.5mm - \hskip -1.5mm
\unitlength 1.00mm
\linethickness{0.4pt}
\begin{picture}(28.00,13.00)(1.00,7.20)
\put(10.00,4.00){\circle*{1.00}}
\put(19.00,4.00){\circle*{1.00}}
\put(19.00,13.00){\circle*{1.00}}
\put(10.00,4.00){\line(1,0){9.00}}
\multiput(19.00,13.00)(0.12,-0.35){6}{\line(0,-1){0.35}}
\multiput(19.00,13.00)(-0.12,-0.35){6}{\line(0,-1){0.35}}
\put(19.00,4.00){\line(0,1){9.00}}
\multiput(19.00,4.00)(-0.35,0.12){6}{\line(-1,0){0.35}}
\multiput(19.00,4.00)(-0.35,-0.12){6}{\line(-1,0){0.35}}
\put(20.00,14.00){\makebox(0,0)[lb]{\scriptsize $x$}}
\put(10.00,3.00){\makebox(0,0)[ct]{\scriptsize $x-e_{\m}-e_{\n}$}}
\put(20.00,4.00){\makebox(0,0)[lt]{\scriptsize $x-e_{\n}$}}
\end{picture}
 \Biggr)\Biggr\}=
}
 \disn{16}{
={1\over 4g^2}\sum\limits_{x,\m,\n}{\rm Tr\,}\Biggl(
\unitlength 1.00mm
\linethickness{0.4pt}
\begin{picture}(21.00,15.50)
(2.0,8.00)
\put(10.00,4.00){\circle*{1.00}}
\put(9.00,14.00){\circle*{1.00}}
\put(11.00,12.00){\circle*{1.00}}
\put(19.00,13.00){\circle*{1.00}}
\multiput(10.00,4.00)(-0.11,1.11){9}{\line(0,1){1.11}}
\multiput(9.50,9.00)(0.10,0.35){6}{\line(0,1){0.35}}
\multiput(9.50,9.00)(-0.14,0.35){6}{\line(0,1){0.35}}
\multiput(9.00,14.00)(1.11,-0.11){9}{\line(1,0){1.11}}
\multiput(9.00,14.00)(0.35,0.10){6}{\line(1,0){0.35}}
\multiput(9.00,14.00)(0.35,-0.14){6}{\line(1,0){0.35}}
\put(20.00,14.00){\makebox(0,0)[lb]{\scriptsize $x$}}
\put(10.00,3.00){\makebox(0,0)[ct]{\scriptsize $x-e_{\m}-e_{\n}$}}
\put(10.00,15.50){\makebox(0,0)[cb]{\scriptsize $x-e_{\m}$}}
\multiput(10.00,4.00)(0.11,0.89){9}{\line(0,1){0.89}}
\multiput(10.90,12.00)(0.10,-0.35){6}{\line(0,-1){0.35}}
\multiput(10.90,12.00)(-0.14,-0.35){6}{\line(0,-1){0.35}}
\multiput(11.00,12.00)(0.89,0.11){9}{\line(1,0){0.89}}
\multiput(15.30,12.50)(-0.35,0.10){6}{\line(-1,0){0.35}}
\multiput(15.30,12.50)(-0.35,-0.14){6}{\line(-1,0){0.35}}
\end{picture}
 \hskip -1.5mm +
\unitlength 1.00mm
\linethickness{0.4pt}
\begin{picture}(28.00,15.00)
(2.20,7.20)
\put(19.00,13.01){\circle*{1.00}}
\put(20.00,3.01){\circle*{1.00}}
\put(18.00,5.01){\circle*{1.00}}
\put(10.00,4.01){\circle*{1.00}}
\multiput(19.00,13.01)(0.11,-1.11){9}{\line(0,-1){1.11}}
\multiput(19.50,8.01)(-0.10,-0.35){6}{\line(0,-1){0.35}}
\multiput(19.50,8.01)(0.12,-0.30){7}{\line(0,-1){0.30}}
\multiput(20.00,3.01)(-1.11,0.11){9}{\line(-1,0){1.11}}
\multiput(20.00,3.01)(-0.35,-0.10){6}{\line(-1,0){0.35}}
\multiput(20.00,3.01)(-0.30,0.12){7}{\line(-1,0){0.30}}
\multiput(19.00,13.01)(-0.11,-0.89){9}{\line(0,-1){0.89}}
\multiput(18.10,5.01)(-0.10,0.35){6}{\line(0,1){0.35}}
\multiput(18.10,5.01)(0.11,0.26){8}{\line(0,1){0.26}}
\multiput(18.00,5.01)(-0.89,-0.11){9}{\line(-1,0){0.89}}
\multiput(13.70,4.51)(0.35,-0.10){6}{\line(1,0){0.35}}
\multiput(13.70,4.51)(0.30,0.12){7}{\line(1,0){0.30}}
\put(20.00,14.00){\makebox(0,0)[lb]{\scriptsize $x$}}
\put(10.00,2.50){\makebox(0,0)[ct]{\scriptsize $x-e_{\m}-e_{\n}$}}
\put(21.50,3.00){\makebox(0,0)[lt]{\scriptsize $x-e_{\n}$}}
\end{picture}
 -
\unitlength 1.00mm
\linethickness{0.4pt}
\begin{picture}(28.00,13.00)(1.00,7.20)
\put(10.00,4.00){\circle*{1.00}}
\put(19.00,4.00){\circle*{1.00}}
\put(19.00,13.00){\circle*{1.00}}
\put(10.00,13.00){\circle*{1.00}}
\put(10.00,4.00){\line(1,0){9.00}}
\multiput(19.00,13.00)(0.12,-0.35){6}{\line(0,-1){0.35}}
\multiput(19.00,13.00)(-0.12,-0.35){6}{\line(0,-1){0.35}}
\put(19.00,4.00){\line(0,1){9.00}}
\multiput(19.00,4.00)(-0.35,0.12){6}{\line(-1,0){0.35}}
\multiput(19.00,4.00)(-0.35,-0.12){6}{\line(-1,0){0.35}}
\put(20.00,14.00){\makebox(0,0)[lb]{\scriptsize $x$}}
\put(10.00,3.00){\makebox(0,0)[ct]{\scriptsize $x-e_{\m}-e_{\n}$}}
\put(20.00,4.00){\makebox(0,0)[lt]{\scriptsize $x-e_{\n}$}}
\put(10.00,4.00){\line(0,1){9.00}}
\multiput(10.00,4.00)(0.12,0.35){6}{\line(0,1){0.35}}
\multiput(10.00,4.00)(-0.12,0.35){6}{\line(0,1){0.35}}
\put(10.00,13.00){\line(1,0){9.00}}
\multiput(10.00,13.00)(0.35,0.12){6}{\line(1,0){0.35}}
\multiput(10.00,13.00)(0.35,-0.12){6}{\line(1,0){0.35}}
\put(10.00,14.50){\makebox(0,0)[cb]{\scriptsize $x-e_{\m}$}}
\end{picture}
 \hskip -1.5mm -
\unitlength 1.00mm
\linethickness{0.4pt}
\begin{picture}(28.00,13.00)(1.00,7.20)
\put(10.00,4.00){\circle*{1.00}}
\put(19.00,4.00){\circle*{1.00}}
\put(19.00,13.00){\circle*{1.00}}
\put(10.00,13.00){\circle*{1.00}}
\put(10.00,4.00){\line(1,0){9.00}}
\multiput(19.00,4.00)(0.12,0.35){6}{\line(0,1){0.35}}
\multiput(19.00,4.00)(-0.12,0.35){6}{\line(0,1){0.35}}
\put(19.00,4.00){\line(0,1){9.00}}
\multiput(10.00,4.00)(0.35,0.12){6}{\line(1,0){0.35}}
\multiput(10.00,4.00)(0.35,-0.12){6}{\line(1,0){0.35}}
\put(20.00,14.00){\makebox(0,0)[lb]{\scriptsize $x$}}
\put(10.00,3.00){\makebox(0,0)[ct]{\scriptsize $x-e_{\m}-e_{\n}$}}
\put(20.00,4.00){\makebox(0,0)[lt]{\scriptsize $x-e_{\n}$}}
\put(10.00,4.00){\line(0,1){9.00}}
\multiput(10.00,13.00)(0.12,-0.35){6}{\line(0,-1){0.35}}
\multiput(10.00,13.00)(-0.12,-0.35){6}{\line(0,-1){0.35}}
\put(10.00,13.00){\line(1,0){9.00}}
\multiput(19.00,13.00)(-0.35,0.12){6}{\line(-1,0){0.35}}
\multiput(19.00,13.00)(-0.35,-0.12){6}{\line(-1,0){0.35}}
\put(10.00,14.30){\makebox(0,0)[cb]{\scriptsize $x-e_{\m}$}}
\end{picture}
\Biggr)
\nom}
One can see that this action is real and nonnegative.

In the limit $a\to 0$ the quantities $\tilde{D}_{\m}\f$, $G_{\m\n}$,
and $S_1$ become
 \disn{17}{
\tilde{D}_{\m}\f(x)\str{a\to 0}
D_{\m}\f(x)+gB_{\m}(x)\f(x),
\nom}
 \disn{18}{
G_{\m\n}(x)\str{a\to 0} iF_{\m\n}(x)=D_{\m}B_{\n}
-D_{\n}B_{\m}+g\lks B_{\m}, B_{\n} \rks,
\nom}
 \disn{19}{
S_1 \str{a\to 0} {1\over 4}\int d^4x \sum\limits_{\m,\n}{\rm Tr}\left\{
F^2_{\m\n}+{\ls D_{\m} B_{\n}-
D_{\n} B_{\m}\rs}^2-\right.\no
\left. -g^2{\rm Tr}\ls {\lks B_{\m},
B_{\n} \rks}^2 \rs-4i F_{\m\n} \lks B_{\m},
B_{\n} \rks \right\},
\nom}
where
 \disn{20}{
F_{\m\n}= \dd_{\m} A_{\n}-\dd_{\n} A_{\m}+
ig \lks A_{\m}, A_{\n} \rks,
\nom}
 \disn{21}{
D_{\m} B_{\n}= \dd_{\m} B_{\n}+
ig \lks A_{\m}, B_{\n} \rks.
\nom}
Here we denote by $D_{\m}$ the usual $U(N)$ covariant derivative.
The form of gauge transformation in the limit $a \to 0$ becomes
 \disn{22a}{
i A^{\prime}_{\m}(x)=u(x)i A_{\m}(x)u^+(x)+
{1 \over g}u(x) \dd_{\m}u^+(x),
\nom}
 \disn{22b}{
B^{\prime}_{\m}(x)=u(x) B_{\m}(x)u^+(x).
\nom}
Thus if we switch off the extra field
$B_{\m}={1 \over \sqrt N}Ib_{\m}+\tilde B_{\m}$
in the limit $a \to 0$ we obtain usual
continuous $U(N)$ gauge theory for the field $A_{\m}$.

   There is another way to construct the  action  with  the  same
properties in the limit $a \to 0$. Let us define the  quantities
$H_{\m\n}(x)$ as
 \disn{r7}{
H_{\m\n}(x)={1\over a^2g}\Biggl(
\unitlength 1.00mm
\linethickness{0.4pt}
\begin{picture}(28.00,13.00)(1.00,7.20)
\put(10.00,4.00){\circle*{1.00}}
\put(19.00,4.00){\circle*{1.00}}
\put(19.00,13.00){\circle*{1.00}}
\put(10.00,13.00){\circle*{1.00}}
\put(10.00,4.00){\line(1,0){9.00}}
\multiput(10.00,4.00)(0.35,0.12){6}{\line(1,0){0.35}}
\multiput(10.00,4.00)(0.35,-0.12){6}{\line(1,0){0.35}}
\put(20.00,14.00){\makebox(0,0)[lb]{\scriptsize $x$}}
\put(10.00,3.00){\makebox(0,0)[ct]{\scriptsize $x-e_{\m}-e_{\n}$}}
\put(20.00,4.00){\makebox(0,0)[lt]{\scriptsize $x-e_{\n}$}}
\put(10.00,4.00){\line(0,1){9.00}}
\multiput(10.00,13.00)(0.12,-0.35){6}{\line(0,-1){0.35}}
\multiput(10.00,13.00)(-0.12,-0.35){6}{\line(0,-1){0.35}}
\put(10.00,14.30){\makebox(0,0)[cb]{\scriptsize $x-e_{\m}$}}
\end{picture}
 -
\unitlength 1.00mm
\linethickness{0.4pt}
\begin{picture}(28.00,13.00)(1.00,7.20)
\put(19.00,4.00){\circle*{1.00}}
\put(19.00,13.00){\circle*{1.00}}
\put(10.00,13.00){\circle*{1.00}}
\put(19.00,4.00){\line(0,1){9.00}}
\multiput(19.00,13.00)(0.12,-0.35){6}{\line(0,-1){0.35}}
\multiput(19.00,13.00)(-0.12,-0.35){6}{\line(0,-1){0.35}}
\put(20.00,14.00){\makebox(0,0)[lb]{\scriptsize $x$}}
\put(20.00,4.00){\makebox(0,0)[lt]{\scriptsize $x-e_{\n}$}}
\put(10.00,13.00){\line(1,0){9.00}}
\multiput(10.00,13.00)(0.35,0.12){6}{\line(1,0){0.35}}
\multiput(10.00,13.00)(0.35,-0.12){6}{\line(1,0){0.35}}
\put(10.00,14.50){\makebox(0,0)[cb]{\scriptsize $x-e_{\m}$}}
\end{picture}
 \Biggr),
\quad\hbox{at $\m\ne \n$},\no
H_{\m\m}(x)={1\over a^2g}\Biggl\{\Biggl(\hskip -3mm
\unitlength 1.00mm
\linethickness{0.4pt}
\begin{picture}(20.00,8.03)
(2.00,5.00)
\put(10.00,7.03){\circle*{1.00}}
\put(10.00,5.03){\circle*{1.00}}
\multiput(13.22,6.72)(-0.35,0.14){6}{\line(-1,0){0.35}}
\multiput(13.22,6.72)(-0.35,-0.10){6}{\line(-1,0){0.35}}
\put(19.00,6.03){\circle*{1.00}}
\put(20.00,7.03){\makebox(0,0)[lb]{\scriptsize $x$}}
\put(10.00,8.33){\makebox(0,0)[cb]{\scriptsize $x-e_{\m}$}}
\multiput(9.98,5.02)(1.00,0.12){9}{\line(1,0){1.00}}
\multiput(19.02,6.09)(-1.13,0.12){8}{\line(-1,0){1.13}}
\multiput(13.00,5.40)(0.35,-0.10){6}{\line(1,0){0.35}}
\multiput(13.00,5.40)(0.35,0.14){6}{\line(1,0){0.35}}
\end{picture}
 \Biggr) -
\Biggl(\hskip 2.7mm
\unitlength 1.00mm
\linethickness{0.4pt}
\begin{picture}(23.65,8.39)
(2.00,5.00)
\put(23.20,7.28){\makebox(0,0)[cb]{\scriptsize $x$}}
\put(14.27,8.70){\makebox(0,0)[cb]{\scriptsize $x-e_{\m}$}}
\put(14.17,5.15){\circle*{1.00}}
\put(14.17,7.15){\circle*{1.00}}
\multiput(10.95,5.46)(0.30,-0.10){7}{\line(1,0){0.30}}
\multiput(10.95,5.46)(0.35,0.12){6}{\line(1,0){0.35}}
\put(5.17,6.15){\circle*{1.00}}
\multiput(14.19,7.16)(-1.00,-0.12){9}{\line(-1,0){1.00}}
\multiput(5.15,6.09)(1.13,-0.12){8}{\line(1,0){1.13}}
\multiput(10.09,6.66)(-0.35,0.12){6}{\line(-1,0){0.35}}
\multiput(10.09,6.66)(-0.35,-0.12){6}{\line(-1,0){0.35}}
\put(23.15,6.08){\circle*{1.00}}
\put(5.23,3.10){\makebox(0,0)[cb]{\scriptsize $x-2e_{\m}$}}
\end{picture}
 \Biggr)\Biggr\}.
\nom}
It means that
 \disn{23}{
H_{\m\n}(x)={1\over a^2g} \ls W(x-e_{\m},\n )W^+(x-e_{\n},\m )-
W^+(x, \m)W(x,\n)\rs=\no
=\tilde{\dd}_{\m}V_{\n}+\tilde{\dd}_{\n}V_{\m}^+-
g\ls V^+_{\m}(x)V_{\n}(x)-V_{\n}(x-e_{\m})V^+_{\m}(x-e_{\n})\rs,
\nom}
and $H_{\m\n}(x)=H^+_{\n\m}(x)$.
Under $U(N)$ gauge transformations we get
 \disn{24}{
H'_{\m\n}(x)=U(x-e_{\m})H_{\m\n}(x)U^+(x-e_{\n}).
\nom}
Hence, the quantity
 \disn{25}{
S_2={1\over 4g^2}\sum\limits_{x,\m,\n}{\rm Tr}\ls H^+_{\m\n}(x)H_{\m\n}(x)\rs
\nom}
is lattice gauge invariant and nonnegative. Any linear combination of
quantities $S_1$ and $S_2$ can be used. In particular,
we can find in the limit $a\to 0$ that
 \disn{26}{
{1\over 2}(S_1\bigl|_{a\to 0}+S_2\bigl|_{a\to 0})=\int d^4x\sum\limits_{\m,\n}
{\rm Tr}\left\{{1\over 4}F^2_{\m\n}+{1\over 2}
(D_{\m}B_{\n})^2-{1\over 4}g^2\ls\lks
B_{\m},B_{\n}\rks\rs^2\right\}.
\nom}

In order to switch off the field $B_{\m}$
in the limit $a\to 0$ one  can  add
to the action $U(N)$ the gauge invariant quantity
 \disn{r8}{
S_m={m^2a^2\over 8g^2}\sum\limits_{x,\m}{\rm Tr}\Biggl(
\Bigl(\hskip -3mm
\unitlength 1.00mm
\linethickness{0.4pt}
\begin{picture}(19.50,8.79)
(2.00,5.00)
\put(19.06,8.79){\makebox(0,0)[cb]{\scriptsize $x$}}
\put(10.06,8.76){\makebox(0,0)[cb]{\scriptsize $x-e_{\m}$}}
\put(19.00,5.15){\circle*{1.00}}
\put(19.00,7.15){\circle*{1.00}}
\multiput(15.78,5.46)(0.30,-0.10){7}{\line(1,0){0.30}}
\multiput(15.78,5.46)(0.35,0.12){6}{\line(1,0){0.35}}
\put(10.00,6.15){\circle*{1.00}}
\multiput(19.02,7.16)(-1.00,-0.12){9}{\line(-1,0){1.00}}
\multiput(9.98,6.09)(1.13,-0.12){8}{\line(1,0){1.13}}
\multiput(14.92,6.66)(-0.30,0.10){7}{\line(-1,0){0.30}}
\multiput(14.92,6.66)(-0.35,-0.12){6}{\line(-1,0){0.35}}
\end{picture}
 - I\Bigl)^2\Biggl).
\nom}
It can be written explicitly as follows:
 \disn{27}{
S_m={m^2a^2\over 8g^2}\sum\limits_{x,\m}{\rm Tr}
\left\{(W^+(x,\m)W(x,\m)-I)^2\right\}=\no
={m^2a^2\over 8g^2}\sum\limits_{x,\m}{\rm Tr}
\left\{(V^+_{\m}(x)+V_{\m}(x))^2-2ag(V^+_{\m}(x)+V_{\m}(x))V^+_{\m}(x)V_{\m}(x)\right. +\no
+\left. a^2g^2(V^+_{\m}(x)V_{\m}(x))^2\right\},
\nom}
where the $m$ is some mass parameter.

In the limit $a\to 0$ the $S_m$ becomes:
 \disn{28}{
S_m\str{a\to 0}{m^2\over 2}\int d^4x \sum\limits_{\m}{\rm Tr}
\ls B^2_{\m}\rs={m^2\over 2}\int d^4x\sum\limits_{\m}
\ls b^2_{\m}+{\rm Tr}\ls \tilde B^2_{\m}\rs\rs.
\nom}
Otherside the quantity
 \disn{29}{
S_{m_b}={m_b^2a^2\over 8Ng^2}\sum\limits_{x,\m}
\Biggl( N-{\rm Tr}\Bigl(\hskip -3.3mm
\unitlength 1.00mm
\linethickness{0.4pt}
\begin{picture}(19.50,9.89)
(2.00,5.00)
\put(19.06,8.90){\makebox(0,0)[cb]{\scriptsize $x$}}
\put(10.06,8.90){\makebox(0,0)[cb]{\scriptsize $x-e_{\m}$}}
\multiput(13.58,7.91)(0.35,-0.12){6}{\line(1,0){0.35}}
\multiput(13.58,7.91)(0.35,0.12){6}{\line(1,0){0.35}}
\multiput(15.60,4.30)(-0.35,0.12){6}{\line(-1,0){0.35}}
\multiput(15.60,4.30)(-0.35,-0.12){6}{\line(-1,0){0.35}}
\put(19.00,6.15){\circle*{1.00}}
\put(10.00,6.15){\circle*{1.00}}
\multiput(10.01,6.12)(0.17,0.11){11}{\line(1,0){0.17}}
\multiput(11.88,7.36)(0.37,0.12){5}{\line(1,0){0.37}}
\put(13.76,7.95){\line(1,0){1.87}}
\multiput(15.63,7.89)(0.22,-0.12){15}{\line(1,0){0.22}}
\multiput(9.98,6.15)(0.17,-0.11){11}{\line(1,0){0.17}}
\multiput(11.86,4.91)(0.38,-0.12){5}{\line(1,0){0.38}}
\put(13.75,4.32){\line(1,0){1.88}}
\multiput(15.63,4.39)(0.23,0.12){15}{\line(1,0){0.23}}
\end{picture}
 \Bigr)\Biggr)^2=\no
={m_b^2a^4\over 8N}\sum\limits_{x,\m}\left\{
{\rm Tr}\ls V^+_{\m}(x)+V_{\m}(x)\rs -ag{\rm Tr}\ls V^+_{\m}(x)V_{\m}(x)\rs\right\}^2
\nom}
is gauge invariant and
 \disn{30}{
S_{m_b}\str{a\to 0}{m_b^2\over 2}\int d^4x\sum\limits_{\m}b^2_{\m},
\nom}
where the $m_b$ is another mass parameter. One  can  choose  this
mass parameters to depend on $a$ so that they go to infinity when
$a\to 0$ giving infinite masses to extra fields.

   This leads us to following form of the action on the lattice :
 \disn{31}{
S=cS_1+(1-c)S_2+S_m+S_{m_b},
\nom}
where the $c$ is arbitrary number in the interval $0\le c\le 1$.
For the concretness we consider further only the action of the form
 \disn{32a}{
S=S_1+S_m,
\nom}

  In order to get $SU(N)$ theory it is  necessary  to  add  to  the
action a term keeping only $SU(N)$ gauge invariance  and  switching
off the abelian field $a_{\m}$ in the limit $a\to 0$.
An example of such term is
 \disn{32b}{
S_{det}={m_{det}^2 a^4\over 2Ng^2}\sum\limits_{x,\m}
\left\{\ls \det W^+(x,\m)-1\rs\ls \det W(x,\m)-1\rs\right\}.
\nom}

 In the limit $a\to 0$ we obtain
 \disn{33}{
S_{det}\str{a\to 0}{m_{det}^2\over 2} \int d^4x \sum\limits_{\m}
\ls a_{\m}^2(x)+b_{\m}^2(x)\rs.
\nom}

   We assume that $m_{det}\to \infty$ at $a\to 0$ to switch off
the  field  $a_{\m}$  in  the
limit. For $SU(3)$ theory this term is of sixth order in the fields
$A_{\m}$
and $B_{\m}$ so that we get rather complicated theory.

   We can easily generalize this approach to include "matter" fields localized
at the sites of the lattice (for example the fermion fields). To do this
we can use
for the part of action containing matter fields in
the same form as one on Wilson-Polyakov lattice with the substitution
of variables $W$ and $W^+$ for the corresponding unitary matrix variables.
All complications connected with fermions on the lattice remain in this
approach. In this paper we do not discuss more detaily the theory with
fermion fields.

   Due to compactness of U(N) and SU(N) gauge groups it is possible to use
lattice theory in nonperturbative calculations without any gauge fixing.
This is true despite the noncompactness of the space of our dynamical
variables.
Nevertheless one can fix the gauge. It is not difficult to prove that
by gauge transformation one
can reduce the field $V_{\m}(x)=B_{\m}(x)+iA_{\m}(x)$ to a form where
 \disn{35}{
A_0(x)=0,\quad \forall x
\nom}
(or $A_1(x)=0$, or $A_2(x)=0$, or $A_3(x)=0$).
The part $B_0(x)$ of the field cannot be made equal to
zero simultanously with the $A_0(x)$ by unitary gauge
transformation. Only in the limit $a\to 0$
when extra fields switch off one gets
usual theory in the gauge $A_0=0$.

   The considered lattice model can be used also for
invariant  ultraviolet  regularization  of    perturbation
theory. As usual one can consider  the  corresponding  functional
integral and separate the quadratic  part  of  the  action.  This
"free" part of the  action  is  invariant  with  respect  to  the
abelian analog of gauge transformations  (\reff{1a}), (\reff{1b}).
Such  abelian
group is noncompact. Therefore it is necessary to fix  the  gauge
(for example,  like  in  eq.~(\reff{35}))  when  perturbation  theory  is
applied. This perturbation theory uses  the  propagators  of  the
fields $A_{\mu}$ and $B_{\mu}$ and  the  vertices  contained in  the
nonquadratic part of the action. The number of vertices is finite
and dos not grows when the order of perturbation theory increases
in contrast with perturbation  theory  based  on  Wilson-Polyakov
lattice.

   Fourier transforms of the fields on the lattice  are  periodic
functions of  momenta.  It  is  possible  to  get  different  but
equivalent forms of perturbation theory in momentum  space  using
different localizations of fields on the lattice.  In  particular
we related the field $V_{\mu}(x)$ on the link connecting the  points
$x-e_{\m}$ and $x$ to the point $x$.
In the construction  of  perturbation
theory it can be more convinient to localize this  field  in  the
point $x-(1/2) e_{\m}$. This leads to the modification of the form of propagators
and  vertices  in  momentum  space  although  both  variants   of
perturbation theory are equivalent.

\section{The Hamitonian formulation in Lorentz coordinates with a lattice
in 3-dimensional space}

   Here we consider the lattice only in 3-dimensional space.
The time coordinate $x_0$ remains continuous.  Starting  with  the  $U(N)$
theory we use as before the matrices
$W(x,i)$, $V_i(x)$, $G_{ik}(x)$, ($i=1, 2, 3$).  The  time
component $A_0(x) = A^{\tilde a}_0(x)T^{\tilde a}$,
($\tilde a=0,1,\dots ,N^2-1$) is taken as  in  usual $U(N)$  theory,  i.e.
without the $B_0(x)$ complement. This field is localized  in  the
sites of 3-dimensional lattice.

  We define the covariant derivative $D_0$ as follows:
 \disn{36}{
D_0\f(x)=(\dd_0+igA_0(x))\f(x)
\nom}
and derive the components $G_{0i}(x)$ of tensor fields from the equality
 \disn{37}{
(D_0\tilde D_i-\tilde D_iD_0)\f(x)=gG_{0i}(x)\f(x-e_i).
\nom}
We get
 \disn{38}{
G_{0i}(x)=\dd_0V_i(x)-i\tilde\dd_iA_0(x)+ig\ls A_0(x)V_i(x)-V_i(x)A_0(x-e_i)\rs,\no
G_{i0}(x)=-G_{0i}(x).
\nom}

   The analog of the action $S_1$ considered before  is
 \disn{39}{
S_1={a^3\over 4}\sum_{\vec x}\int dx^0{\rm Tr}\ls 2\sum_i G^+_{0i}(x)G_{oi}(x)-
\sum_{i,k}G^+_{ik}(x)G_{ik}\rs\equiv \int dx^0 L_1,
\nom}
where  $x=x^1,x^2,x^3$ and the minus before the second term is connected
with the transition to pseudoeucledean space. In order to switch off the
extra fields $B_i(x)$ in the limit $a\to 0$ we add to the $S_1$ the
analog of the "mass"  term $S_m$ introduced before:
 \disn{40}{
S_m=-{m^2a^3\over 8}\sum^3_{\vec x,i=1}\int dx^0{\rm Tr}
\left\{ \ls V^+_i(x)+V_i(x)\rs^2+\right. \no
\left.-2ag\ls V^+_i(x)+V_i(x)\rs V^+_i(x)V_i(x)+a^2g^2\ls V^+_i(x)V_i(x)\rs^2\right\}\equiv\int dx^0L_m.
\nom}
The action now is
 \disn{41}{
S=S_1+S_m\equiv\int dx^0L,
\nom}
where $L=L_1+L_m$ . The mass parameter $m$ should be infinitely inceased
when $a\to 0$ in order to switch off the fields $B_i(x)$ .

  For the $SU(N)$ theory we add to the action the analog of the "mass" term
$S_{det}$ introduced before:
 \disn{42}{
S_{det}=-{m^2_{det}a^3\over 2Ng^2}\sum_{\vec x,i}\int dx^0
\left\{ (\det W^+(x,i)-1)(\det W(x,i)-1)\right\}=\int dx^0L_{det},
\nom}
The parameter $m_{det}$ also tends to infinity when $a\to 0$ .

   The transition to the Hamiltonian formulation can  be  carried
out as usual. We introduce at $x^0=const$ the "momenta"
 \disn{43}{
\Pi_i^{\tilde a+}(\vec x)\equiv {\dd L\over\dd\ls\dd_0V_i^{\tilde a}(\vec x)\rs}=
{\dd L_1\over\dd\ls\dd_0V_i^{\tilde a}(\vec x)\rs}={a^3\over 2}G_{0i}^{\tilde a+}(\vec x),
\nom}
conjugated to the $V_i^{\tilde a}(\vec x)$, and the "momenta"
 \disn{44}{
\Pi^{\tilde a}_i(\vec x)\equiv {\dd L\over\dd\ls\dd_0V_i^{\tilde a+}(\vec x)\rs}=
{a^3\over 2}G_{0i}^{\tilde a+}(\vec x),
\nom}
conjugated to the $V_i^{\tilde a+}$. Besides, we get
 \disn{45}{
\pi^{\tilde a}_0(\vec x)\equiv {\dd L\over \dd\ls \dd_0A^{\tilde a}_0(\vec x)\rs}=0.
\nom}
The generalized Hamiltonian is
 \disn{46}{
\tilde H=\sum_{\vec x,i}{\rm Tr}\ls \Pi^+_i(\vec x)\dd_0V_i(\vec x)+
\Pi_i(\vec x)\dd_0V_i^+(\vec x)\rs-L=H-\sum_{\vec x}\ls A_0(\vec x)\phi(\vec x)\rs,
\nom}
where
 \disn{47}{
H=\sum_{\xx}{\rm Tr}\left\{{2\over a^3}\sum_i\Pi^+_i(\xx)\Pi_i(\xx)+
{a^3\over 4}G^+_{ik}(\xx)G_{ik}(\xx)\right\}-L_m,
\nom}
 \disn{48}{
\phi(\xx)=\sum_i\left\{i\ls\tilde \dd_i\Pi^+_i(\xx+\vec e_i)-\tilde\dd_i\Pi_i(\xx+\vec e_i)\rs+\right.\no
\left. +ig\ls V^+_i(\xx+\vec e_i)\Pi_i(\xx+\vec e_i)-\Pi^+_i(\xx+\vec e_i)V_i(\xx+\vec e_i)-
\Pi_i(\xx)V^+_i(\xx)+V_i(\xx)\Pi^+_i(\xx)\rs\right\}.
\nom}
The $\phi(x)$ is the 1st class constraint. The Lagrangian can be written in
Hamiltonian form
 \disn{49}{
L^{(1)}=\sum_{\xx,i}{\rm Tr}\ls\Pi^+_i(\xx)\dd_0V_i(\xx)+\Pi_i(\xx)\dd_0V^+_i(\xx)\rs-
H+\sum_{\xx}{\rm Tr}\ls\tilde A_0(\xx)\phi(\xx)\rs.
\nom}

   For the $SU(N)$ theory the term $(-L_{det})$ is to be added to the $H$.
Separating hermitian and antihermitian parts of the fields
 \disn{50}{
V_i=B_i+iA_i,\qquad \Pi_i={1\over 2}\ls P_i+i\pi_i\rs,
\nom}
where $B^+_i=B_i$, $A^+_i=A_i$, $P^+_i=P_i$, $\pi^+_i=\pi_i$,
we get the pairs of real canonical variables
$(A_i^{\tilde a},\pi_i^{\tilde a})$ and $(B_i^{\tilde a},P_i^{\tilde a})$.

The transition to quantum theory can be realized by two different ways. One
of them is to fix the gauge by the relation $A_0=0$,
limit the physical subspace of states by the condition
 \disn{52}{
\phi(\xx)\left|\Psi_{ph}\right>=0,
\nom}
and to solve the Schroedinger equation
 \disn{53}{
(H-E)\left|\Psi_{ph}\right>=0
\nom}
in the physical subspace. Another way is
to introduce the gauge $A_3=0$, to solve
explicitly the constraint with respect to $\pi_3$ and to substitute it for
the $\pi_3$ in the Hamiltonian. However the presence of nonzero  $B_3$
when $A_3=0$ makes the problem of finding the $\pi_3$ from the constraint
equation practically nonsolvable. Nevertheless one can slightly modify the
model under consideration to get $A_3=B_3=0$
without destroying the gauge
invariance. To do this we relate to the links which are
directed along the $x^3$ axis the unitary matrices
$U(x,3)=\exp(-iag\tilde A_3(x))$
instead of $W(x,3)$ and
the  $U^+(x,3)=\exp(iag\tilde A_3(x))$ instead of $W^+(x,3)$.
Other link variables remain unchanged. The gauge
invariant action can be obtained from previos one by the following changings:
 \disn{56}{
V_3(x)\to {U(x,3)-I\over (-ag)},\qquad V^+_3(x)\to {U^+(x,3)-I\over (-ag)}.
\nom}
In this modified theory it is possible to fix the gauge by the relation
 \disn{57}{
U(x,3)=I,
\nom}
i.~e. $A_3(x)=0$.  After  this  gauge  fixing  the  modified   action
coincides with the action considered before  with  $A_3=B_3=0$.  It
follows that the condition $A_3=B_3=0$ in  the  unmodified  model
agrees in fact with gauge invariance. Using  this  condition  and
looking at the Hamiltonian formalism we see that
 \disn{58}{
P_3=0,\qquad \Pi_3=i\pi_3.
\nom}

The Hamiltonian coincides with the  eq.~(\reff{47})  at  $A_3=B_3=0$  and
$\Pi_3=i\pi_3$. The constraint (\reff{48}) can be  now  easily  resolved
with respect to the $\pi_3$ and the  Hamiltonian can be  written  in
terms of remaining independent variables. It may be noticed  that
one  could  consider  the  model in the  gauge  $A_3=0$ with
continuous $x^0$ and $x^3$ and with  the  lattice  in  the  $x^1$,
$x^2$. Then it would be possible to put  $B_0=B_3=0$  immedeately.
However the ultraviolet  regularization  in  such  model  is  not
complete.

\section{Canonical formulation of gauge theory on the LF}
\subsection{Transverse space lattice with continuous $x^+$ and $x^-$}

As before we use the following denotations for the LF coordinates:
 \disn{4.1}{
x^{\pm}={1\over \sqrt{2}}(x^0\pm x^3),\qquad x^\p\equiv (x^1,x^2)
\equiv x^k,\quad k=1,2,
\nom}
where the $x^+$ plays the role of time \cite{dir}.
Transverse coordinates $x^k$
correspond to the sites of the transverse lattice.

Our canonical formulation of gauge theory on the LF is similar to Bardeen-
Pearson one \cite{bapir} but with using of other independent variables and with more
detailed taking
into account of the $x^-$ zero mode problem \cite{fra,kal}. This problem
can be formulated
canonically on the interval $-L\le  x^-\le   L$
using the assumption of periodic
boundary conditions in the  $x^-$ for gauge fields. Such regularization
preserves translation and gauge ivariance \cite{fra,kal}.

In continuous space this formulation deals with very complicated  2nd class
constraints containing zero modes of gauge fields, and it is not clear how to
treat them in terms of quantum operator variables. We show that the
introduction of transverse
lattice in the framework of such formulation allows to avoid this 2nd class
constraints at all. Furthermore we pesent the consideration of
the LF 1st class consraints \cite{fra,kal}.

   Let us start with the $U(N)$ theory of pure gauge fields. The
components of fields corresponding to continuous coordinates $x^+,x^-$ can
be taken in a form
 \disn{4.2}{
V_{\pm}=iA_{\pm},\qquad A_{\pm}^+=A_{\pm}
\nom}
because this simplification is allowed by a form of gauge transformations.
Then we get
in the same way as the eqv.~(\reff{38}) was derived the folowing relations:
 \disn{4.3}{
G_{+-}(x)=iF_{+-}(x),\no G_{\pm k}(x)=\dd_{\pm}V_k(x)-i\tilde \dd_kA_{\pm}(x)
+ig\ls A_+(x)V_k(x)-V_k(x)A_+(x-e_k)\rs.
\nom}

To write the action in canonical (Hamiltonian) form we fix the gauge
appropriately to periodic boundary
conditions in the $x^-$ \cite{fra}:
 \disn{4.4}{
\dd_- A_- =0,\qquad A_-^{ij}(x)=\de^{ij} v^j(x^+,x^{\p}).
\nom}

The $i,j$ are the $U(N)$ matrix indeces ($i,j = 1,2,...,N $).
We obtain \vskip -7mm
 \disn{4.5}{
S={a^2\over 2}\sum\limits_{x^\p}\int dx^+\int\limits_{-L}^Ldx^-
{\rm Tr}\ls G_{+-}^+G_{+-}+\sum_{k=1,2}\ls
G_{+k}^+G_{-k}+G_{-k}^+G_{+k}\rs-G_{12}^+G_{12}\rs_x+S_{m\p}=\no
={a^2\over 2}\sum_{x^{\p}}\int dx^+\int\limits_{-L}^Ldx^-
\ls{\rm Tr}\lks 2G_{+-}^+G_{+-}+\sum_{k=1,2}\ls
G_{+k}^+G_{-k}+G_{-k}^+G_{+k}\rs \rks-H\rs_x=\no
={a^2\over 2}\sum_{x^{\p}}\int dx^+\int\limits_{-L}^Ldx^-
\left\{\sum_{i=1}^N2L\lks 2(F_{+-})_x+i\tilde\dd_k(V_k-V^+_k)_{x+e_k}\rks^{ii}_{(0)} \dd_+v^i(x)\right.+\no
+\sum_{i,j=1}^N\lks\ls (\dd_--igv^i(x)+igv^j(x-e_k))V^{+ij}_k(x)\rs
\dd_+V^{ij}_k(x)+h.c.\rks+\no
\left. +\sum_{i,j=1}^N A_+^{ij}(x)Q^{ji}(x)-H(x)\right\},
\nom} \vskip -3mm \noindent
where the denotation $()_x$ means that all quantities inside the brackets
are to be taken at the point $x$; the $S_{m\p}$ is the
"mass" term given by eq. (\reff{27}) (with opposite sign) where one takes into account
the transition to the transverse lattice (then it does
depend only on transverse components of the fields); furthermore
 \disn{4.6}{
f_{(0)}\equiv {1\over 2L}\int\limits_{-L}^L dx^-f(x^-),
\nom}
and the $H(x)$ is the Hamiltonian density.

  The generators of gauge transformations $Q^{ij}(x)$ are defined by the
following expression
 \disn{4.7}{
Q(x)=2(D_-F_{+-})_x+i\tilde \dd_k(G^+_{-k}-G_{-k})_{x+e_k}+\no
+ig(V_kG^+_{-k}-G_{-k}V^+_k)_x-
ig(G^+_{-k}V_k-V^+_kG_{-k})_{x+e_k}.
\nom}

The gauge constraints
 \disn{4.8}{
Q^{ij}(x)=0
\nom}
can be resolved explicitly by expressing the quantities $F^{ij}_{+-}$
in terms of other
variables except of zero mode components $F^{ii}_{+-(0)}$
which can not be found from this
constraint equation. So the corresponding zero mode
$Q^{ii}_{(0)}(x^{\p},x^+)$ of the constraint
remains unresolved explicitly and it is considered as the condition on the
physical quantum states:
 \disn{4.9}{
Q^{ii}_{(0)}(x^\p,x^+)\left|\Psi_{phys}\right>=0.
\nom}

   In order to complete the derivation of the action in canonical form and
to extract all independent canonical variables we make the Fourier
transformation in the $x^-$ of the transverse field components
$V^{ij}_k(x)$ as follows:
 \disn{4.10}{
V^{ij}_k(x)=\sum_{n=-\infty }^{\infty }\left\{ \Theta \ls p_n-gv^i(x)+
gv^j(x-e_k)\rs V^{ij}_{nk}(x^\p,x^+)+\right. \no
\left. +\Theta \ls -p_n+gv^i(x)-
gv^j(x-e_k)\rs V^{ij+}_{nk}(x^\p,x^+) \right \} \times \no
\times \ls 4L \left | p_n-gv^i(x)+
gv^j(x-e_k) \right | \rs ^{-1/2} e^{-ip_nx^-},
\nom}
where $p_n=\pi n/L, n \epsilon Z$.

   Then the action is
 \disn{4.11}{
S={a^2 \over 2}\sum_{x^{\p}} \int dx^+ \left\{ \sum_{i=1}^N 2L \lks 2(F_{+-})_x+
i\tilde {\dd_k}(V_k-V_k^+)_{x+e_k} \rks^{ii}_{(0)} \dd_+v^i(x)+\right. \no
\left. +i\sum_{n=-\infty }^{\infty }\sum_{i,j=1}^N \ls V^{ij+}_{nk}
\dd _+V^{ij}_{nk}\rs_x +2L \sum_{i=1}^N \ls A_{+(0)}^{ii}Q_{(0)}^{ii} \rs_x-
\bar H(x) \right\},
\nom}
where the quantities $\ls V_k \rs^{ii}_{(0)}$ are expressed in terms
of variables $V^{ii}_{0k}$ according to eqs. (\reff{4.6}), (\reff{4.10})
as follows:
 \disn{4.12}{
\ls V_k \rs^{ii}_{(0)}={V^{ii}_{0k} \over \sqrt{4Lga|\tilde{\dd _k}v^i|}}
\nom}
and the $\bar H$ is obtained from the $H$ via substitution of
the expressions for the $F_{+-}^{ij}$ found from the constraints
(\reff{4.8}).

   We have the following set of canonically conjugated pairs of independent
variables:
 \disn{4.13}{
\left \{ v^i(x),\qquad \Pi _i(x)=La^2\lks 2(F_{+-})_x+
i \tilde{\dd _k}(V_k-V^+_k)_{x+e_k} \rks^{ii}_{(0)}\right \}, \no
\left \{ V^{ij}_{nk}(x),\qquad {ia^2 \over 2}V^{ij+}_{nk}(x) \right \}.
\nom}

   In quantum theory this variables become operators which satisfy usual
canonical commutation relations.

   In the obtained formulation there are no 2nd class constraints for zero
modes of the transverse field components. If one goes to the limit $a \to 0$
this constraints reappear in a form which contains quantum operators in
definite order. This order was not clear earlier.

 One can easily construct canonical operator of translations in the $x^-$ :
 \disn{4.14}{
P_-^{can.}={a^2 \over 2} \sum_{x^{\p}} \sum_{k=1,2} \sum_n \sum_{i,j=1}^N
p_n \e \ls p_n-gv^i(x)+gv^j(x-e_k) \rs \ls V_{nk}^{ij+}V_{nk}^{ij} \rs_x.
\nom}

This expression differs from the physical gauge invariant momentum
operator $P_- $ by a term proportional to the consraint. The operator $P_-$ is
 \disn{4.15}{
P_-=a^2 \sum_{x^{\p}} \sum_{k=1,2} \int\limits_{-L}^L dx^- {\rm Tr} \ls G^+_{-k}
G_{-k} \rs _x=
P_-^{can.} + La^2  \sum_{x^{\p}} \sum_{i=1}^N \ls v^iQ^{ii}_{(0)} \rs _x = \no
={a^2 \over 2} \sum_{x^{\p}} \sum_{k=1,2} \lks ga \sum_{i=1}^N \ls V^{ii}_{0k}-
\sqrt{{4L \over ga} | \tilde{\dd_k}v^i|} \rs^+_x \ls V^{ii}_{0k}-
\sqrt{{4L \over ga} | \tilde{\dd_k}v^i|} \rs_x+\right. \no
\left. +{\sum_{n,i,j}}' \left| p_n-gv^i(x)+gv^j(x-e_k) \right|
\ls V^{ij+}_{nk}V^{ij}_{nk} \rs_x \rks
\nom}
where the ${\sum}'$ denotes the sum over all $n, i, j$ except of $i=j$ at $n=0$.

   The operators $Q^{ii}_{(0)}(x^{\p},x^+)$  have the following form in
terms of canonical variables:
 \disn{4.16}{
2LQ^{ii}_{(0)}(x)=\sum_{k=1,2} \left\{ -ga \tilde{\dd_k} \ls \e (\tilde{\dd_k}v^i)
\ls V^{ii}_{0k}-\sqrt{{4L \over ga} | \tilde{\dd_k}v^i|}\rs^+
\ls V^{ii}_{0k}-\sqrt{{4L \over ga} | \tilde{\dd_k}v^i|}\rs \rs_{x+e_k}+\right. \no
+g {\sum_{n,j}}'\lks \e (p_n-gv^j(x+e_k)-gv^i(x))\ls V^{ji+}_{nk}V^{ji}_{nk}\rs_{x+e_k}-\right.\no
\left.\left.-
\e (p_n-gv^i(x)+gv^j(x-e_k)) \ls V^{ij+}_{nk}V^{ij}_{nk}\rs_x \rks \right\}.
\nom}

   In order to find general form of physical states satisfying the condition
(\reff{4.9}) it is convinient to intrduce a basis consisting of following
state vectors:
 \disn{4.17}{
\prod_{x^{\p}}\prod_{k=1,2}\prod_n \prod_{i,j=1}^N
\ls V^{ij}_{nk}(x)\rs^{m^{ij}_{nk}(x)}
\ls R^i_k(x) \rs^{m^i_k(x)} \left| \bar v \right>
\nom}
where we use "creation and annihilation" operators,
$R_k^{i+}(x)$ and $R^i_k(x)$, defined as follows :
 \disn{4.18}{
R^i_k(x)\equiv\ls V^{ii}_{0k}-\sqrt{{4L \over ga} | \tilde{\dd_k}v^i|}\rs_x
\nom}

      The $R^i_k(x)$ act on the vector $\left| \bar v \right>$
like annihilation operators:
 \disn{4.19}{
R^i_k(x)\left| \bar v \right>=0
\nom}
The vector $\left| \bar v \right>$ is connected with the standart vector
$\left|v \right>$ satisfyine the conditions
 \disn{4.20}{
\hat v^i(x)\left|v\right>=v^i(x)\left|v\right>,
\nom}
 \disn{4.21}{
V^{ij}_{nk}(x)\left|v\right>\equiv 0,
\nom}
by the following transformation:
 \disn{4.22}{
\left|\bar v\right>=\exp\left\{ {a^2\over 2} \sum_{x^{\p}}\sum_{k=1,2}\sum_{i=1}^N
\ls{4L\over ga}|\tilde{\dd_k}v^i|\rs_x^{1/2}V^{ii+}_{0k}(x)\right\}\left|v\right> .
\nom}

    A set of nonnegative integer numbers $m^{ij}_{nk}(x)$ is defined so that
 \disn{4.23}{
m^{ii}_{0k}(x)\equiv 0.
\nom}

   In this basis the conditions (\reff{4.9}) take the form of following
relations between the numbers $m^{ij}_{nk}(x), m^i_k(x)$ and the $v^i(x)$:
 \disn{4.24}{
\sum_{k=1,2}\left\{ -a\tilde{\dd_k}\ls\e (\tilde{\dd_k}v^i)m^i_k\rs_{x+e_k}+
\sum_n\sum_{j=1}^N\lks\e (p_n-gv^j(x+e_k)+gv^i(x))m^{ij}_{nk}(x+e_k)-
\right.\right.\no
\left. \left. -\e (p_n-gv^i(x)+gv^j(x-e_k))m^{ij}_{nk}(x)\rks\right\}=0
\nom}

   One can find the eigenvalue $p_-$ of the momentum $P_-$ for such basis
state:
 \disn{4.25}{
p_-=\sum_{x^{\p}}\sum_{k=1,2}\lks ga\sum_{i=1}^N\ls |\tilde{\dd_k}v^i|m^i_k\rs_x+
\sum_n\sum_{i,j=1}^N|p_n-gv^i(x)+gv^j(x-e_k)|m^{ij}_{nk}(x)\rks
\nom}
and to require that this value be finite. The detailed analysis of this
problem will be given in future publication.

  Let us discuss the transition to the $SU(N)$ gauge theory.
In order to avoid abelian part of the field in the limit
$a\to 0$ we add to the action the additional "mass term" which
is obtained from the expression (\reff{32b})
with opposite sign
by transition to transverse
lattice. Moreover we can restrict the general form of the  $A_+, A_-$
by the condition
 \disn{4.26}{
{\rm Tr}A_+={\rm Tr}A_-=0.
\nom}

    We get in this way the theory with the same canonical structure in transverse
variables as before. According to eq. (\reff{4.26}) one has to choose the independent
components among $v^i$ and $\Pi_i$, and to modify the expression for
the constraint operator $Q^{ij}(x)$ by subtraction of the
$(N^{-1}){\rm Tr}Q(x)$. Analogous modification has to be done in the
formulation of the problem (\reff{4.9}).

\subsection{Lattice in coordinates $x^1$, $x^2$ and $x^-$}

    The  formulation  of  gauge  theory  on  the  LF  with
transverse lattice and continuous coordinates  $x^+$, $x^-$
gives only partial ultraviolet regularization.
To complete the regularization we can introduce the lattice  also
along the $x^-$. Let us discuss this possibility. We consider for
simplicity the U(N) theory in the space with
 unbounded coordinate $x^-$ .
   We introduce  different  lattice  spacing  parameters  in  the
transverse and in  the  $x^-$  coordinates.  Let  us  denote  the
transverse one as before by $a$ and the one along the $x^-$ direction by
$b$. The $x^+$ coordinate remains  continuous, and  corresponding
field component is taken in the simplest  form  $V_+=iA_+$,  i.e.
$B_+=0$. We relate the matrix $W(x, -)$ to the link  directed  to
positive side of the $x^-$ axis , and the  matrix  $W^+(x,-)$
to the link directed to the opposite side. The  matrix  $W(x,-)$
is related to the link connecting the point $x-e_-$ with  the
point $x$. We put
 \disn{4b.1}{
W(x,-)=I-gbV_-(x),\quad V_-(x)=B_-(x)+iA_-(x).
\nom}
The component $G_{+-}$ is defined as follows
 \disn{4b.2}{
G_{+-}(x)=\dd_+V_-(x)-i\tilde\dd_-A_+(x)+ig\ls A_+(x)V_-(x)-V_-(x)A_+(x-e_-)\rs.
\nom}
Let us take into account that the components $G_{+k}(x)$,
$(k=1,2)$ change under
gauge transformation as follows
 \disn{4b.3}{
G'_{+k}(x)=u(x)G_{+k}(x)u^+(x-e_k),
\nom}
and choose the definition of the $G_{-k}(x)$ so that the quantity
 \disn{4b.4}{
{\rm Tr}\ls G^+_{+k}(x)G_{-k}(x)\rs,
\nom}
entering into the action remains gauge invariant. Such definition is
 \disn{4b.5}{
G_{-k}(x)={1\over 2gab}\Biggl(2
\unitlength 1.00mm
\linethickness{0.4pt}
\begin{picture}(25.00,17.31)
(2.00,10.20)
\put(10.00,7.01){\circle*{1.00}}
\put(19.00,7.01){\circle*{1.00}}
\put(19.00,16.01){\circle*{1.00}}
\put(10.00,16.01){\circle*{1.00}}
\put(10.00,7.01){\line(1,0){9.00}}
\multiput(10.00,7.01)(0.35,0.12){6}{\line(1,0){0.35}}
\multiput(10.00,7.01)(0.35,-0.12){6}{\line(1,0){0.35}}
\put(20.00,17.01){\makebox(0,0)[lb]{\scriptsize $x$}}
\put(10.00,6.01){\makebox(0,0)[ct]{\scriptsize $x-e_--e_k$}}
\put(20.00,7.01){\makebox(0,0)[lt]{\scriptsize $x-e_k$}}
\put(10.00,7.01){\line(0,1){9.00}}
\multiput(10.00,13.00)(0.12,-0.35){6}{\line(0,-1){0.35}}
\multiput(10.00,13.00)(-0.12,-0.35){6}{\line(0,-1){0.35}}
\put(10.00,17.31){\makebox(0,0)[cb]{\scriptsize $x-e_-$}}
\put(10.00,15.99){\line(1,0){9.00}}
\multiput(19.00,16.02)(-0.30,-0.10){7}{\line(-1,0){0.30}}
\multiput(19.00,16.02)(-0.35,0.12){6}{\line(-1,0){0.35}}
\end{picture}
 -
\unitlength 1.00mm
\linethickness{0.4pt}
\begin{picture}(22.00,16.12)
(-3.00,9.00)
\put(16.00,14.60){\makebox(0,0)[lc]{\scriptsize $x$}}
\put(14.17,13.62){\circle*{1.00}}
\put(14.17,15.62){\circle*{1.00}}
\multiput(10.95,13.93)(0.30,-0.10){7}{\line(1,0){0.30}}
\multiput(10.95,13.93)(0.35,0.12){6}{\line(1,0){0.35}}
\put(5.17,14.62){\circle*{1.00}}
\multiput(14.19,15.63)(-1.00,-0.12){9}{\line(-1,0){1.00}}
\multiput(5.15,14.56)(1.13,-0.12){8}{\line(1,0){1.13}}
\multiput(10.09,15.13)(-0.35,0.12){6}{\line(-1,0){0.35}}
\multiput(10.09,15.13)(-0.30,-0.10){7}{\line(-1,0){0.30}}
\put(5.92,16.01){\makebox(0,0)[rb]{\scriptsize $x-e_-$}}
\multiput(14.17,13.54)(0.10,-0.30){7}{\line(0,-1){0.30}}
\multiput(14.17,13.54)(-0.10,-0.30){7}{\line(0,-1){0.30}}
\put(14.17,5.83){\circle*{1.00}}
\put(14.19,5.83){\line(0,1){7.79}}
\put(14.27,4.73){\makebox(0,0)[ct]{\scriptsize $x-e_k$}}
\end{picture}
-
\unitlength 1.00mm
\linethickness{0.4pt}
\begin{picture}(27.00,16.19)
(-2.50,9.00)
\put(16.00,6.82){\makebox(0,0)[lc]{\scriptsize $x-e_k$}}
\put(14.17,5.84){\circle*{1.00}}
\put(14.17,7.84){\circle*{1.00}}
\multiput(10.95,6.15)(0.35,-0.12){6}{\line(1,0){0.35}}
\multiput(10.95,6.15)(0.35,0.12){6}{\line(1,0){0.35}}
\put(5.17,6.84){\circle*{1.00}}
\multiput(14.19,7.85)(-1.00,-0.12){9}{\line(-1,0){1.00}}
\multiput(5.15,6.78)(1.13,-0.12){8}{\line(1,0){1.13}}
\multiput(10.09,7.35)(-0.30,0.10){7}{\line(-1,0){0.30}}
\multiput(10.09,7.35)(-0.30,-0.10){7}{\line(-1,0){0.30}}
\put(5.92,4.81){\makebox(0,0)[ct]{\scriptsize $x-e_--e_k$}}
\multiput(14.17,15.55)(0.10,-0.30){7}{\line(0,-1){0.30}}
\multiput(14.17,15.55)(-0.10,-0.30){7}{\line(0,-1){0.30}}
\put(14.17,15.69){\circle*{1.00}}
\put(14.19,7.84){\line(0,1){7.79}}
\put(16.09,15.65){\makebox(0,0)[lc]{\scriptsize $x$}}
\end{picture}
 \Biggr)=\no
={1\over 2gab}\left\{ 2W(x,-)W(x-e_-,k)W^+(x-e_k,-)- \right.\no
\left. -W(x,-)W^+(x,-)W(x,k)-W(x,k)W(x-e_k,-)W^+(x-e_k,-)\right\}.
\nom}
This expression agrees in the continuous limit with the usual tensor field
if the extra fields $B_k$, $B_-$ are
switched off in this limit. Now the action can be written in the form
 \disn{4b.6}{
S_1={a^2b\over 2}\sum_{x^-,x^1,x^2}\int dx^+{\rm Tr}
\left\{ G^+_{+-}(x)G_{+-}(x)+\right.\no
\left.+G^+_{+k}(x)G_{-k}(x)+G^+_{-k}(x)G_{+k}(x)-G^+_{12}(x)G_{12}(x)\right\}.
\nom}
We add also necessary "mass" terms in order to switch off the extra fields
in the continuous limit.

  In order to get the canonical formalism we have to fix the gauge $A_-=0$.
However the $B_-$ part cannot be simultanously set equal to zero
by gauge transformation. This leads to the difficulty with  solving of the
canonical constraints
 \disn{4b.7}{
\pi_k^+=G^+_{-k},
\nom}
where $\pi_k^+$ is the momentum
conjugate to $A_k$.
To pass over this difficulty we again apply the
modified model in which the $W(x, -)$ is rechanged by unitary matrix
$U(x,-)=\exp\ls -igbA_-(x)\rs $.
The components $G_{12}, G_{+k}$ remain unchanged. The quantities
$G_{-k}$ become as follows:
 \disn{4b.8}{
G_{-k}(x)={1\over gab}\Biggl(\hskip 1.5mm
\unitlength 1.00mm
\linethickness{0.4pt}
\begin{picture}(25.00,17.31)
(2.00,10.20)
\put(10.00,7.01){\circle*{1.00}}
\put(19.00,7.01){\circle*{1.00}}
\put(19.00,16.01){\circle*{1.00}}
\put(10.00,16.01){\circle*{1.00}}
\put(10.00,7.01){\line(1,0){9.00}}
\multiput(10.00,7.01)(0.35,0.12){6}{\line(1,0){0.35}}
\multiput(10.00,7.01)(0.35,-0.12){6}{\line(1,0){0.35}}
\put(20.00,17.01){\makebox(0,0)[lb]{\scriptsize $x$}}
\put(10.00,6.01){\makebox(0,0)[ct]{\scriptsize $x-e_--e_k$}}
\put(20.00,7.01){\makebox(0,0)[lt]{\scriptsize $x-e_k$}}
\put(10.00,7.01){\line(0,1){9.00}}
\multiput(10.00,13.00)(0.12,-0.35){6}{\line(0,-1){0.35}}
\multiput(10.00,13.00)(-0.12,-0.35){6}{\line(0,-1){0.35}}
\put(10.00,17.31){\makebox(0,0)[cb]{\scriptsize $x-e_-$}}
\put(10.00,15.99){\line(1,0){9.00}}
\multiput(19.00,16.02)(-0.30,-0.10){7}{\line(-1,0){0.30}}
\multiput(19.00,16.02)(-0.35,0.12){6}{\line(-1,0){0.35}}
\end{picture}
 -
\unitlength 1.00mm
\linethickness{0.4pt}
\begin{picture}(20.00,17.01)
(10.00,10.20)
\put(19.00,7.01){\circle*{1.00}}
\put(19.00,16.01){\circle*{1.00}}
\put(19.00,7.01){\line(0,1){9.00}}
\multiput(19.00,13.00)(0.12,-0.35){6}{\line(0,-1){0.35}}
\multiput(19.00,13.00)(-0.12,-0.35){6}{\line(0,-1){0.35}}
\put(20.00,17.01){\makebox(0,0)[lb]{\scriptsize $x$}}
\put(20.00,7.01){\makebox(0,0)[lt]{\scriptsize $x-e_k$}}
\end{picture}
 \Biggr)=\no
={1\over gab}\ls U(x,-)W(x-e_-,k)U^+(x-e_k,-)-W(x,k)\rs, \no
G_{+-}(x)=\dd_+\ls {U(x,-)-I\over (-ag)}\rs -i\tilde\dd_- A_+(x)+\no
+ig\ls A_+(x)\ls{U(x,-)-I\over (-ag)}\rs -\ls{U(x,-)-I\over (-ag)}\rs A_+(x-e_-)\rs.
\nom}
Further we substitute this expressions for the $G_{\mu\nu}$ in the action
 \disn{4b.9}{
S=S_1+S_m
\nom}
where $S_1$ is defined  by eq. (\ref{4b.6})  and $S_m$
is the same as in eq. (\ref{4.5}).
We get again $U(N)$ gauge invariant action which has correct
naive continuous limit if $m\to \infty$ when $a\to 0$, $b\to 0$.
In the modified theory one can use the gauge
 \disn{4b.10}{
A_-=0,
\nom}
i.e.
 \disn{4b.11}{
U(x,-)=I.
\nom}
In this gauge one has
 \disn{4b.12}{
G_{+-}(x)=-i\tilde\dd_- A_+(x), \qquad  G_{-k}(x)=\tilde\dd_-V_k(x),
\nom}
and there remain previous expressions for the $G_{+k}, G_{12}$.
Now there is no difficulty  with  solving the canonical
constraints, and all steps necessary for the construction of Hamiltonian
formalism can be done in standard way.

   Let us discuss the question connected with the regularization of the
"infrared" LF
divergences at $p_-=0$. To achieve this regularization one can as usual
take the lattice with finite
number of sites in the $x^-$ direction and choose periodic boundary
conditions in the $x^-$. However this requiers gauge fixing in the form
(as in the section 4.1)
 \disn{4b.14}{
A_-(x)=v(x^+,x^1,x^2),
\nom}
where the $v$ is a diagonal matrix. In this gauge the unitary matrices
$U(x, -)$ become diagonal but the action remains nonpolynomial in variables
$v$. More detailed investigation of this question is under consideration.

\section*{Acknowledgements}
This work was in part supported by the DFG  Grant
436 RUS 113/205/1 on Russian-German cooperation. We thank B. Van de Sande
for interesting discussions.


\begin{thebibliography}{99}

\bibitem{vilpol}
K.~G.~Wilson. Phys.~Rev. D10 (1974) 2445.

\bibitem{bapir}
W.~A.~Bardeen, R.~B.~Pearson. Phys.~Rev. D14 (1976) 547. W.~A.~Bardeen,
R.~B.~Pearson, E.~Rabinovici. Phys.~Rev. D21 (1980) 1037.

\bibitem{dal1}
S.~Dalley, B.~van de Sande. Nucl.~Phys. B53 (Proc.~Suppl.) (1997) 827.

\bibitem{dal2}
S.~Dalley, B.~van de Sande. Hep-ph/9704408.

\bibitem{fra}
V.~A.~Franke, Yu.~V.~Novozhilov, E.~V.~Prokhvatilov.
Lett.~Math.~Phys. 5 (1981) 239,437.

\bibitem{kal}
A.~C.~Kalloniatis, H.~C.~Pauli. Z.~Phys. C 60 (1993) 255, C 63 (1994) 161,
A.~C.~Kalloniatis. Phys.~Rev. D 54 (1996) 2876, H.~C.~Pauli,
A.~C.~Kalloniatis, S.~S.~Pinsky. Phys.~Rev. D 52 (1995) 1176.

\bibitem{dir}
P.~A.~M.~Dirac. Rev.~Mod.~Phys. 81 (1949) 392.

\end{thebibliography}
\end{document}